\documentclass[12pt]{revtex4}
\usepackage{amsmath}
\usepackage[dvips]{graphicx}
\usepackage{dcolumn}
\usepackage{bm}
\usepackage{epsfig}
\usepackage{color}

\input{tcilatex}
\begin{document}

\title{Dynamics of end to end loop formation for an isolated chain in
viscoelastic fluid}
\author{Rajarshi Chakrabarti}

\begin{abstract}
We theoretically investigate the looping dynamics of a linear polymer immersed in a viscoelastic fluid. The dynamics of the chain is governed by a Rouse model with a fractional memory kernel recently proposed by Weber et al. (S. C. Weber, J. A. Theriot, and A. J. Spakowitz, Phys. Rev. E 82, 011913 (2010)). Using the Wilemski-Fixman (G. Wilemski and M. Fixman, J. Chem. Phys. 60, 866 (1974)) formalism we calculate the looping time for a chain in a viscoelastic fluid  where the mean square displacement of the center of mass of the chain scales as $t^{1/2}$. We observe that the looping time is faster for the chain in viscoelastic fluid than for a Rouse chain in Newtonian fluid up to a chain length and above this chain length the trend is reversed. Also no scaling of the looping time with the length of the chain seems to exist for the chain in viscoelastic fluid.
\end{abstract}

\affiliation{Institute for Computational Physics, University of Stuttgart, Pfaffenwaldring 27 70569, Stuttgart, Germany}
\maketitle

\section{Introduction}

Looping dynamics involving long chain molecules has been an active research
topic in chemical physics \cite{wilemski1974, doi, szabo, sebastianjcp,
thirumalai, toan}. Obviously loop formation is a primary step in protein
folding and thus finds lot of attention from chemists and biologists too
\cite{eaton, hudgins}. It has been shown that the hydrodynamic interaction
\cite{friedman, chakrabartiphysica}, flexibility of the chain \cite{santo,
cherayiljcp20021, cherayiljcp20022, hyeon} and the solvent quality \cite%
{cherayiljcp2004} profoundly alters the looping time. But to our best
knowledge the effect of viscoelastic fluid around a
polymer chain on its looping dynamics has not been addressed yet. But it has been shown that the presence of viscoelastic fluid around the
 polymer changes the mode relaxation of polymers and thus it should also influence the looping dynamics. The importance of such
study where a polymer is immersed in a viscoelastic fluid comes from recent
experiments on biopolymers in viscoelastic environment such as cytoplasm \cite{ernstsoftmatter}.
It has been shown very recently that the dynamics of the chromosomal loci in the viscoelastic bacterial cells gets greatly affected and the diffusion becomes anomalous \cite{spakowitzprl}. Diffusion of lipid granules inside viscoelastic cells have also found out to be anomalous \cite{jeonprl2011}.  In this paper we address the problem of end to end loop formation for a chain immersed in a viscoelastic solvent where the chain dynamics is anomalous and is described by a Rouse model with fractional memory kernel \cite{spakowitzpre}. For
this we use the recently proposed model of a single polymer chain in a
viscoelastic fluid \cite{spakowitzpre}. A Similar model have earlier been
used in the context of mode coupling theory of polymeric fluids \cite{schweizer}.

The paper is arranged as follows. In section II the Wilemski-Fixman theory for the end to end loop formation is briefly discussed. The radial delta function sink used in the calculation is introduced in section III. Section IV deals with the model for a polymer in viscoelastic fluid. Section V presents the results and the paper ends with the conclusion in section VI.

\section{Theory of end to end loop formation: Wilemski-Fixman formalism}

In this section we briefly discuss the Wilemski-Fixman formalism for the end to end loop formation in long chains. The dynamics of a single polymer chain with reactive end-groups is modeled by the following Smoluchowski equation \cite{doibook, kawakatsubook}.

\begin{equation}
\frac{\partial P(\{\mathbf{R}\},t)}{\partial t}=LP(\{\mathbf{R}\},t)-kS(\{%
\mathbf{R}\})P(\{\mathbf{R}\},t)  \label{Smolu1}
\end{equation}

Here $P(\{R\},t)$ is the distribution function for the chain that it has the
conformation $\{R\}\equiv R_{1},$$R_{2},$$......R_{n}$ at time $t$ where $%
R_{i}$ denotes the position of the $i$th monomer in the chain of $n$
monomers. $S({R})$ is called the sink function which actually models the
reaction between the ends and thus usually is a function of end to end
vector. $L$ is a differential operator, defined as

\begin{equation}
L=D_{0}\sum_{i=1}^{n}\frac{\partial }{\partial \mathbf{R}_{i}}.\left[ \frac{%
\partial }{\partial \mathbf{R}_{i}}+\frac{\partial U}{\partial \mathbf{R}_{i}%
}\right] P(\{\mathbf{R}\},t)  \label{operator}
\end{equation}

Here $D_0$ is the diffusion coefficient of the chain defines as the inverse
of the friction coefficient per unit length and $U$ is the potential energy
of the chain. Wilemski and Fixman \cite{wilemski1974} then derived an approximate expression for
the mean first passage time from Eq. (\ref{Smolu1}). This mean first passage
time is actually the loop closing time for the chain. Thus only valid in the
limit of infinite sink strength, $k\to\infty$ \cite{RednerBook,
chakrabartijcp}. The expression for this loop closing time reads

\begin{equation}
\tau =\int_{0}^{\infty }dt\left( \frac{C(t)}{C(\infty )}-1\right)
\label{tau}
\end{equation}

\noindent Where $C(t)$ is the sink-sink correlation function defined as

\begin{equation}
C(t)=\int d\mathbf{R}\int d\mathbf{R}_{0}S(\mathbf{R})G(\mathbf{R},t|\mathbf{%
R}_{0},0)S(\mathbf{R}_{0})P(\mathbf{R}_{0})  \label{ct}
\end{equation}

\noindent In the above expression $G(\mathbf{R},t|\mathbf{R}_{0},0)$ is the
the conditional probability that a chain with end-to-end distance $R_0$ at
time $t=0$ has the end-to-end distance $R$ at time $t$; $P(R_0)$ is the
equilibrium distribution of the end to end distance since the chain is in
equilibrium at time $t=0$. $S(R)$ is the sink function \cite%
{sebastianpra1992, debnath, chakrabarticpl2010} which depends only on the
separation between the chain ends.

Now it is obvious that the knowledge of $G(\mathbf{R},t| \mathbf{R}_{0},0)$
is prerequisite to calculate the sink-sink correlation function $C(t)$ and
hence the closing time $\tau$. In case of a flexible chain the Greens
function and the end-to-end probability distribution functions are known and
Gaussian.

For a flexible chain the Greens function is given by

\begin{equation}
G(\mathbf{R},t|\mathbf{R}_{0},0)=\left( \frac{3}{2\pi \left\langle
R^{2}\right\rangle _{eq}}\right) ^{3/2}\frac{1}{\left( 1-\phi ^{2}(t)\right)
^{3/2}}\times \exp \left[ -\frac{3(\mathbf{R}-\phi (t)\mathbf{R}_{0})^{2}}{%
2\left\langle R^{2}\right\rangle _{eq}(1-\phi ^{2}(t))}\right]
\label{greens}
\end{equation}

Where

\begin{equation}
\phi (t)=\frac{\left\langle \mathbf{R}(t).\mathbf{R}(0)\right\rangle _{eq}}{%
\left\langle R^{2}\right\rangle _{eq}}  \label{phi}
\end{equation}

is the normalized end-to-end vector correlation function for the chain. The
above ensemble average is taken over the initial equilibrium distribution
for end-to-end vector $P(\mathbf{R}_{0})$.

\bigskip

Similarly end-to-end equilibrium distribution for the flexible chain with $%
L_1=N b^2$ ($b$ is the kuhn length and $N$ is the number of
monomers) at time $t=0$ is given by \cite{doibook, sokolov}

\begin{equation}
P(R_{0})=\left( \frac{3}{2\pi L_1^{2}}\right) ^{3/2}\exp \left[ -\frac{%
3R_{0}^{2}}{2L_1^{2}}\right]  \label{peq}
\end{equation}

With the above Gaussian functions the sink-sink correlation function can be
written as a radial double integral.

\begin{eqnarray}
C(t) &=&\left( \frac{3}{2\pi L_1^{2}}\right) ^{3}\frac{1}{\left( 1-\phi
^{2}(t)\right) ^{3/2}}\int_{0}^{\infty }4\pi R^{2}S(R)dR\int_{0}^{\infty
}4\pi R_{0}^{2}S(R_{0})dR_{0}\times  \notag \\
&&\exp \left[ -\frac{3}{2L_1^{2}}\frac{(R^{2}+R_{0}^{2})}{(1-\phi ^{2}(t))}%
\right] \frac{\sinh \left[ (3\phi (t)RR_{0})/(L_1^{2}(1-\phi ^{2}(t)))\right]
}{(3\phi (t)RR_{0})/(L_1^{2}(1-\phi ^{2}(t)))}  \label{sink-sink}
\end{eqnarray}

The above integral can be evaluated analytically for some specific choice of
the sink functions. With a radial delta function sink the above integral can
be evaluated analytically \cite{pastor}.

\section{The radial delta function sink}

For a radial delta function sink \cite{pastor}, $%
S(R)=\delta (R-a) $, where $a$ is the capture radius in the model the sink-sink correlation function defined in (Eq.(\ref{sink-sink})) can be evaluated exactly. Since in
this case the integration over $R$ and $R_0$ can be carried out
analytically, the looping time can be expressed in a closed form as follows.

\begin{equation}
\tau =\int_{0}^{\infty }dt\left( \frac{\exp \left[ -(2x_{0}\phi
^{2}(t))/(1-\phi ^{2}(t))\right] \sinh \left[ (2x_{0}\phi (t))/(1-\phi
^{2}(t))\right] }{2x_{0}\phi (t)\sqrt{1-\phi ^{2}(t)}}-1\right)
\label{taudelta}
\end{equation}

with
\begin{equation*}
x_{0}=\frac{3a^{2}}{2L_{1}^{2}}
\end{equation*}

Obviously if the end-to-end vector correlation function $\phi (t)$ is known
the closing time can be calculated by carrying out the integration over
time. Throughout the paper it is assumed that the chain is in equilibrium
at $t=0$ and only at $t=0^{+}$ the viscoelasticity is turned on. Thus the
equilibrium end-to-end distribution for the Rouse as well as for the
Rouse chain with the fractional memory kernel is given by (Eq.(\ref{peq})). The last integration
over time is not analytical so has to be carried out numerically.

\section{model of polymer in viscoelastic fluid}

It has been shown recently that a viscoelastic fluid background imparts a memory into the dynamics of the isolated chain immersed in the fluid. So the dynamics of a chain in a viscoelastic
fluid is modeled by a generalized Langevin equation \cite{spakowitzpre}.
In other words the dynamics of an isolated chain without self interaction
and hydrodynamic interaction in a viscoelastic fluid is described by a Rouse model with
fractional Langevin model kernel. Within this model, a linear chain described by a
space curve $R(n,t)$ has the following equation of motion

\begin{equation}
\xi \int\limits_{0}^{t}dt_{1}K(t-t_{1})\frac{\partial \mathbf{R}(n,t)}{%
\partial t}=\frac{3k_{B}T}{b^{2}}\frac{\partial ^{2}\mathbf{R}(n,t)}{%
\partial n^{2}}+\mathbf{f}(n,t).  \label{gle}
\end{equation}

The random force $f(n,t)$ is Gaussian \cite{xie} and satisfies the following fluctuation dissipation theorem \cite{spakowitzpre}

\begin{equation}
\left\langle \mathbf{f}(n,t)\mathbf{f}(n_{1},t_{1})\right\rangle =\xi
K(t-t_{1})\delta (n-n_{1})\mathbf{I.}  \label{fdt}
\end{equation}

\noindent where $\xi$ is the friction coefficient and $K(t-t_1)$ is the memory kernel.

\noindent This is exactly the model adopted by Weber et al. \cite{spakowitzpre} to investigate the physics
of subdiffusive motion of a polymer in viscoelastic fluid. Although practically the same model has earlier
been used by Min et al. \cite{xie} in the context of fluctuation of the distance between a donor-acceptor pair within a single protein complex \cite{minpre2006, chakrabartijcp2007}. Also the Smoluchowski equation (Eq.(\ref{Smolu1})) no longer applies to the above chain, as the chain dynamics is non-Markovian in this case.

\noindent In this case the
normalized end to end vector time correlation function (Eq.(\ref{phi})) is
given by

\begin{equation}
\phi (t)=\frac{\left\langle \mathbf{R}(t).\mathbf{R}(0)\right\rangle _{eq}}{%
\left\langle R^{2}\right\rangle _{eq}}=16\sum\limits_{p=odd}\left\langle
\mathbf{X}_{p}(t)\mathbf{X}_{p}(0)\right\rangle _{eq}  \label{phit}
\end{equation}

where the correlation function has been shown to have the following expression
\cite{spakowitzpre}

\begin{equation}
\left\langle \mathbf{X}_{p}(t)\mathbf{X}_{p}(0)\right\rangle _{eq}=\frac{%
3k_{B}T}{k_{p}}E_{\alpha ,1}\left[ -\frac{k_{p}}{N\xi \Gamma (3-\alpha )}%
t^{\alpha }\right]   \label{xpcorr}
\end{equation}

where $k_p=\frac{6\pi^2 k_B T}{N b^2}p^2$ and $E_{\alpha, \beta}(x)$ is the
generalized Mittag-Leffler function \cite{erdelyibook}

\begin{equation}
E_{\alpha ,\beta }(x)=\sum\limits_{i=0}^{\infty }\frac{x^{i}}{\Gamma (\beta
+\alpha i)}  \label{mlf}
\end{equation}

The generalized Mittag-Leffler function reduces to the regular
Mittag-Leffler function for $\beta =1$. In the limit $\alpha \rightarrow 1$,
which is the case of Rouse chain in Newtonian fluid the regular
Mittag-Leffler function becomes simple exponential and one gets back well
known Rouse relaxation dynamics. But in a viscoelastic fluid for which $%
0<\alpha <1$, $\phi (t)$ obviously decays non-exponentially and one would expect this to profoundly affect
the end to end loop formation. In principle $\phi (t)$ defined in Eq.(\ref{phit}%
) should be put back into Eq.(\ref{taudelta}) to carry out the integration
over time to get the looping time. This would be the end to end looping time
for a linear chain without any self or hydrodynamic interaction in a
viscoelastic fluid. Unfortunately the regular Mittag-Leffler function (Eq.(%
\ref{mlf}) with $\beta =1$) can not be evaluated analytically for an
arbitrary value of $\alpha $.  In general when $\alpha=1/n$ where $n=2,3,4,..$ the
regular Mittag-Leffler function can be expressed as a sum over incomplete gamma functions. We choose $\alpha =1$, which is the case for a
Rouse chain in Newtonian fluid and $\alpha =1/2$ which is a special case for
a Rouse chain in a viscoelastic fluid the regular Mittag-Leffler function
has analytically exact expressions. With $\alpha =1/2$ \cite{xie}, the
regular Mittag-Leffler function becomes $\ E_{1/2,1}(x)=\exp (x^{2})Erfc(-x)$%
, where \ $Erfc(x)=1-\frac{2}{\sqrt{\pi }}\int\limits_{0}^{x}\exp (-y^{2})dy$
is the complimentary error function. Using this exact expression for $\alpha
=1/2$ one arrives at the analytically exact expression for $\phi _{\alpha =1/2}(t)$.

\begin{equation}
\phi _{\alpha =1/2}(t)=\sum\limits_{p=odd}\frac{8}{\pi ^{2}p^{2}}E_{1/2,1}%
\left[ -(t/\tau _{p, \alpha=1/2})^{1/2}\right] =\sum\limits_{p=odd}\frac{8}{\pi
^{2}p^{2}}\exp (t/\tau _{p, \alpha=1/2})Erfc(({t/\tau _{p, \alpha=1/2}})^{1/2})
\label{phitvisco}
\end{equation}

where $\tau _{p, \alpha=1/2}=\frac{\xi^2 N^{4}b^{4}}{64(k_{B}T)^2\pi ^{3}}\frac{1}{p^4}$.

Weber et al. \cite{spakowitzpre} showed that with $\alpha=1/2$ the mean square displacement of the center of mass of the polymer scales as $t^{1/2}$.

\bigskip

On the other hand for a Rouse chain in Newtonian fluid for which $\alpha =1$, the above
expression becomes a sum over exponentials as all the odd modes relaxes
exponentially.

\begin{equation}
\phi _{\alpha =1}(t)=\sum\limits_{p=odd}\frac{8}{\pi ^{2} p^2}\exp
(-t/\tau _{p, \alpha=1})  \label{phitrouse}
\end{equation}

where $\tau _{p, \alpha=1}=\frac{\xi N^{2}b^{2}}{3k_{B}T\pi ^{2}}\frac{1}{p^2}$.

Notice $\phi_{\alpha=1}(t \rightarrow 0)=1$ as  $\sum\limits_{p=odd}\frac{8}{\pi ^{2}p^{2}}=1$ and $\phi_{\alpha=1}(t \rightarrow \infty)=0$. Same is true for $\phi_{\alpha=1/2}$ but it behaves as a stretched exponential at short $t$ and inverse power law at long $t$. This can be seen by expanding
$exp (t/\tau _{p, \alpha=1/2})Erfc(({t/\tau _{p, \alpha=1/2}})^{1/2})$ at short $t$ and doing an asymptotic expansion of the same at large $t$.

\section{results}

The looping time with a radial delta function sink defined in (Eq.(\ref{taudelta})) is calculated for a chain in the viscoelastic fluid when the dynamics of the chain is described by a generalized Langevin equation (Eq.(\ref{gle})). It is done as follows. First the end to end vector time correlation function is calculated for the chain using Eq.(\ref{phitvisco}) as only for this case with $\alpha=1/2$ the correlation function can be evaluated exactly. Then $\phi_{\alpha=1/2}(t)$ is put back in (Eq.(\ref{taudelta})) and the integration over time is carried out numerically to get the looping time. Plots of $\phi_{\alpha=1/2}(t)$ and $\phi_{\alpha=1}(t)$ (for the Rouse chain in Newtonian fluid) are shown in Fig. 1. Notice the same set of parameters are used along with the same value of the friction coefficient ($\xi$).  As one can see that the initial decay of $\phi(t)$ is faster in viscoelastic fluid than in Newtonian but beyond a time scale the trend is reversed and $\phi_{\alpha=1/2}(t)$ approaches zero extremely slow. Also for a shorter chain the initial decay of $\phi_{\alpha=1/2}(t)$ is even faster. Faster the $\phi(t)$  decays faster the integrand in Eq.(\ref{taudelta}). Thus for a short chain immersed in the viscoelastic fluid the contribution from the initial decay is so small that it did not get compensated from the longtime contribution and results a faster looping. While for a long chain the longtime dynamics contributes appreciably so that the area under the integrand is big enough to make loop formation slower.  This can be seen from the plot of $ln(\tau)$ vs $ln(N)$ in Fig. 2. Another important observation is the absence of any well defined scaling of the looping time with the chain length for the chain in viscoelastic fluid ($\alpha=1/2$). This is also evident from the
$ln(\tau)$ vs $ln(N)$ plot in Fig. 2. But the Rouse chain in the Newtonian fluid ($\alpha=1$) shows a $\tau\sim N^2$ scaling as expected \cite{cherayiljcp2004, chakrabartiphysica}.

\begin{figure}[tbp]
\centering
\epsfig{file=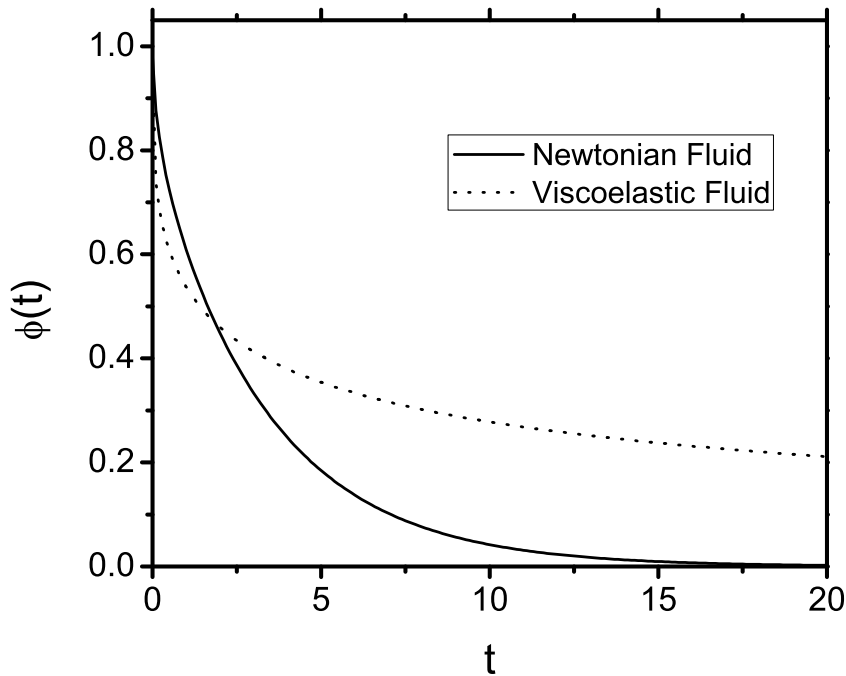,width=0.8\linewidth}\newline
\caption{$\protect\phi(t)$ against time ($t$). The values of
parameters used are: $N=100, b=0.1, \protect\xi=1, k_BT=1 $.}
\label{phicompare}
\end{figure}


\begin{figure}[tbp]
\centering
\epsfig{file=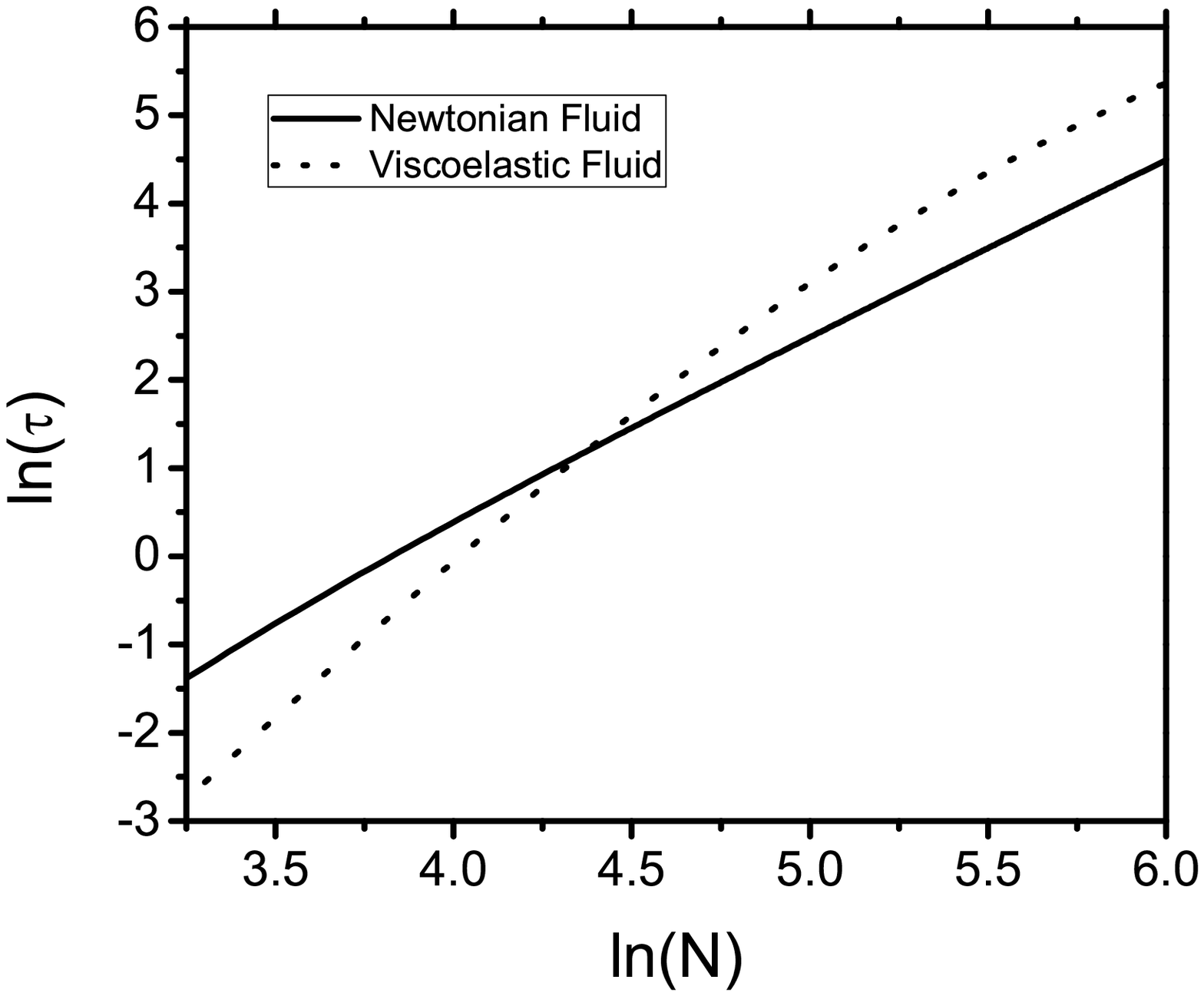,width=0.8\linewidth}\newline
\caption{$ln(\tau)$ against $ln(N)$. The values of
parameters used are: $a=0.5, b=0.1, \protect\xi=1, k_BT=1 $.}
\label{logtau}
\end{figure}

\section{conclusions}

In this paper we calculate the looping time for a chain immersed in a viscoelastic fluid where the mean square displacement of the center of mass of the chain scales as $t^{1/2}$. We found that up to a chain length viscoelastic fluid actually enhances the looping rate as compared to a Newtonian fluid with the same friction. But beyond a chain length the trend is reversed. This observation is due to the faster short time decay of the normal modes of the chain in the viscoelastic fluid as compared to in a Newtonian fluid. This is eve more prominent for short chains. Also no scaling of the looping time with the chain length seems to exist in viscoelastic fluid. Here we would like to point out that in a real viscoelastic fluid a chain would definitely feel a larger frictional drag. Thus in a real viscoelastic fluid the chain would relax even slower and one would expect slower looping time as compared to normal Rouse chain even with short chains.

In future we would like to investigate the effect of self interaction along with the viscoelastic background on the looping dynamics. It is obvious that in a crowded environment self interactions will also have profound influence on the looping dynamics of the chains.

\section{acknowledgement}

This work was carried out during the author's stay at the Indian Institute
of Science and supported by the Department of Science and Technology through
the J. C. Bose fellowship project of K. L. Sebastian. The author thanks K.
L. Sebastian for encouragements.

\bibliographystyle{apsrev}

\end{document}